\documentclass[12pt,preprint]{emulateapj}

\def\be{\begin{equation}}
\def\ee{\end{equation}}
\def\bea{\begin{eqnarray}}
\def\eea{\end{eqnarray}}
\usepackage{graphicx}
\usepackage{amsmath}
\usepackage[colorlinks,
            linkcolor=blue,
            anchorcolor=blue,
            citecolor=blue]{hyperref}

\usepackage{color,txfonts}
\begin{document}

\title{HESS J1427-608: an unusual hard unbroken $\gamma-$ray spectrum in a very wide energy range}

\author{Xiao-Lei Guo\altaffilmark{1,2}, Yu-Liang Xin\altaffilmark{2,3}, Neng-Hui Liao\altaffilmark{2},
Qiang Yuan\altaffilmark{2}, Wei-Hong Gao\altaffilmark{1}, Hao-Ning He\altaffilmark{2}, Yi-Zhong Fan\altaffilmark{2}, Si-Ming Liu\altaffilmark{2}}
\email{E-mail:yuanq@pmo.ac.cn (QY); gaoweihong@njnu.edu.cn (WHG); liusm@pmo.ac.cn (SML)}

\altaffiltext{1}{Department of Physics and Institute of Theoretical Physics, Nanjing Normal University, Nanjing 210046, China}

\altaffiltext{2}{Key laboratory of Dark Matter and Space Astronomy, Purple Mountain Observatory, Chinese Academy of Sciences, Nanjing 210008, China}

\altaffiltext{3}{University of Chinese Academy of Sciences, Yuquan Road 19, Beijing, 100049, China}

\begin{abstract}
We report the detection of a GeV $\gamma$-ray source which is spatially overlapping  and thus very likely associated
with the unidentified very-high-energy (VHE) $\gamma$-ray source HESS J1427-608 
with the Pass 8 data recorded by the Fermi Large Area Telescope. The photon spectrum 
of this source is best described by a power-law with an index of $1.85\pm0.17$ in 
the energy range of $3-500$ GeV, and the measured flux connects smoothly with that 
of HESS J1427-608 at a few hundred GeV. 
This source shows no significant extension and time variation.
The broadband GeV-TeV emission over four decades of 
energies can be well fitted by a single power-law function with an index of 2.0, 
without obvious indication of spectral cutoff toward high energies. Such a result 
implies that HESS J1427-608 may be a PeV particle accelerator. We discuss possible 
nature of HESS J1427-608 according to the multi-wavelength spectral fittings. 
Given the relatively large errors, either a leptonic or a hadronic model can explain 
the multi-wavelength data from radio to VHE $\gamma$-rays. The inferred magnetic field 
strength is a few $\mu$G, which is smaller than typical values of supernova remnants (SNRs), 
and is consistent with some pulsar wind nebulae (PWNe). On the other hand, the flat $\gamma$-ray 
spectrum is slightly different from typical PWNe but similar to that of some known SNRs.

\end{abstract}


\setlength{\parindent}{.25in}

\section{Introduction}

The High Energy Stereoscopic System (HESS) survey of the inner
Galaxy \citep{Aharonian2006a} has discovered a large number of very
high energy (VHE; $>$ 100 GeV) sources. About half of them have
been firmly identified as supernova remnants (SNRs) and pulsar wind
nebulae (PWNe). However, there is still a significant fraction of VHE 
sources that has not been identified due to the lack of counterparts
in the radio/X-ray or GeV $\gamma$-ray band \citep{Aharonian2005,Aharonian2008a}. 
Therefore, searching for the counterparts of these unidentified VHE 
sources in longer wavelength bands is very helpful to identify and 
classify them. In the GeV band, more than seven years survey data 
from the Fermi Large Area Telescope (Fermi-LAT) provides a good 
opportunity to hunt for the unidentified VHE sources. Indeed, several 
unidentified VHE sources have been detected with the Fermi-LAT data \citep{Hui2016}.

HESS J1427-608 is one of the unidentified VHE sources discovered by the HESS 
Galactic Plane Survey \citep{Aharonian2008b}. The TeV image shows that it 
is only slightly extended with a symmetric Gaussian function with width of 
$\sigma$ = $3'$. This source is detected with an integral flux above 1 TeV 
of $F_{\gamma}(>1 {\rm TeV})$ = $4.0 \times 10^{-12} $ erg cm$^{-2}$ s$^{-1}$
and a power-law spectrum with an index of $\Gamma$ = 2.16. No nearby SNR 
or pulsar was detected in the vicinity of HESS J1427-608 \citep{Green2014}. 
The X-ray counterpart of it, Suzaku J1427-6051, was reported by \citet{Fujinaga2013}. 
The X-ray spatial morphology is clearly extended with a Gaussian width of 
$\sigma$ = $0.9' \pm 0.1'$ and the spectrum is well fitted by a power-law with 
an index of $\Gamma$ = $3.1^{+0.6}_{-0.5}$, implying non-thermal 
X-ray continuum emission. The unabsorbed flux in the 2-10 keV band is 
$F_{\rm X}=(9^{+4}_{-2}) \times 10^{-13}~{\rm erg}~{\rm cm}^{-2}~{\rm s}^{-1}$. \citet{Fujinaga2013} 
has discussed the possible nature of HESS J1427-608 as a PWN or non-thermal SNR. 
Each one can explain some of the observational facts but face difficulties.
No radio counterpart was detected in the direction of HESS J1427-608.
In the second Fermi-LAT catalog \citep[2FGL;][]{Nolan2012}, a GeV source,
2FGL J1427.6-6048c, which is located in the vicinity of HESS J1427-608, 
was reported. However, the central position of them differs moderately 
from each other. Furthermore, the power-law spectrum index of 2FGL J1427.6-6048c 
is $\Gamma$ = 2.7 \citep{Nolan2012}, which can not connect with the TeV spectrum 
of HESS J1427-608 well. It is very likely that 2FGL J1427.6-6048c is not associated 
with HESS J1427-608. Meanwhile, it should be noted that no source in 
the third Fermi-LAT source catalog \citep[3FGL;][]{Acero2015} is associated with 2FGL J1427.6-6048c.

In this paper, we carry out a complete analysis of this region using
more than 7 years Fermi-LAT Pass 8 data to investigate the
$\gamma$-ray emission of HESS J1427-608. In section 2, we describe the data
analysis and results, including the spatial, spectral and timing analysis.
The possible nature of HESS J1427-608 based on the multi-wavelength
spectral energy distribution (SED) fitting is discussed in section 3 
and the conclusion of this work is presented in section 4.

\section{Data analysis}

\subsection{Data reduction}

\begin{figure*}[!htb]
\centering
\includegraphics[width=3.5in]{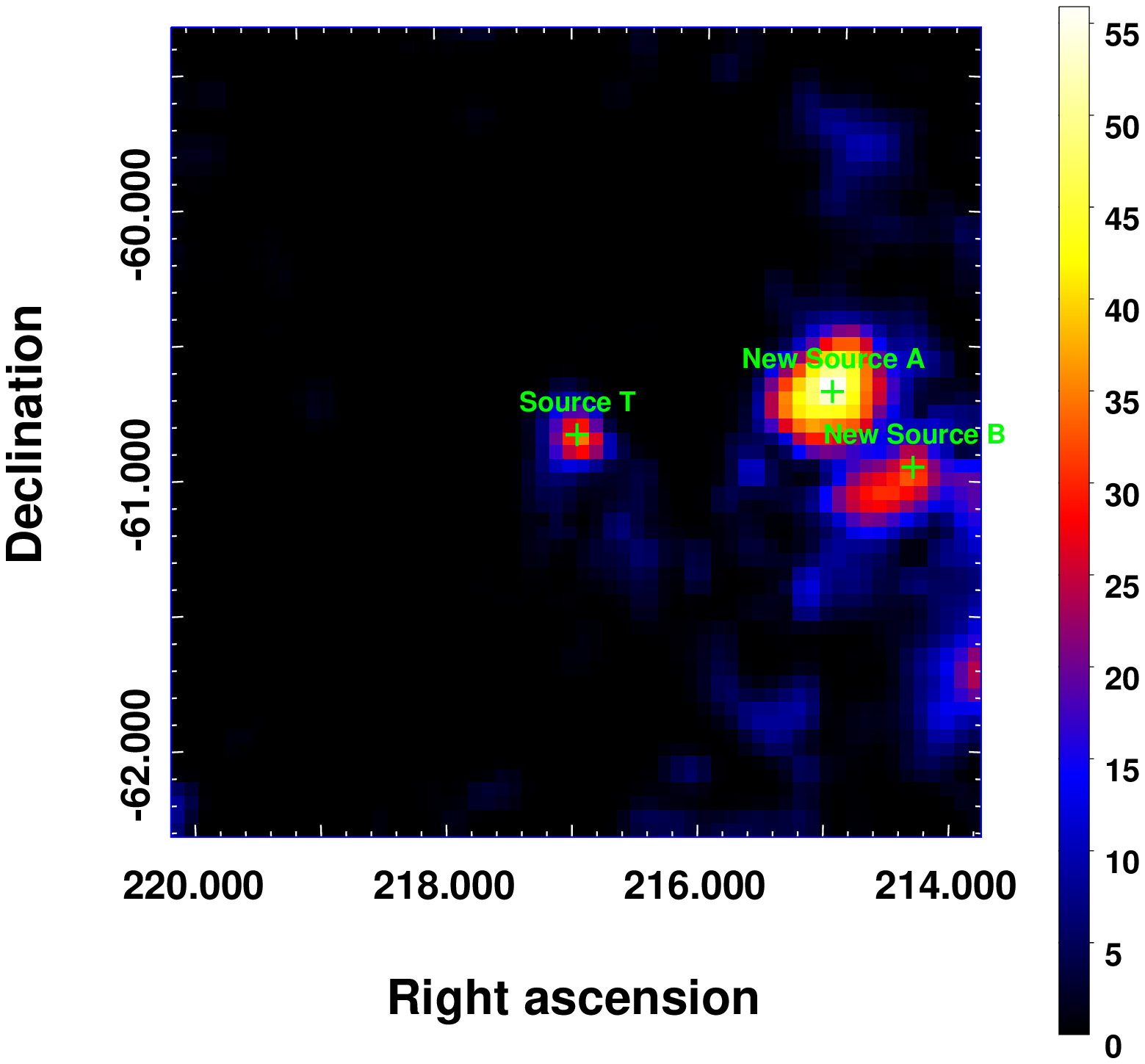}
\includegraphics[width=3.5in]{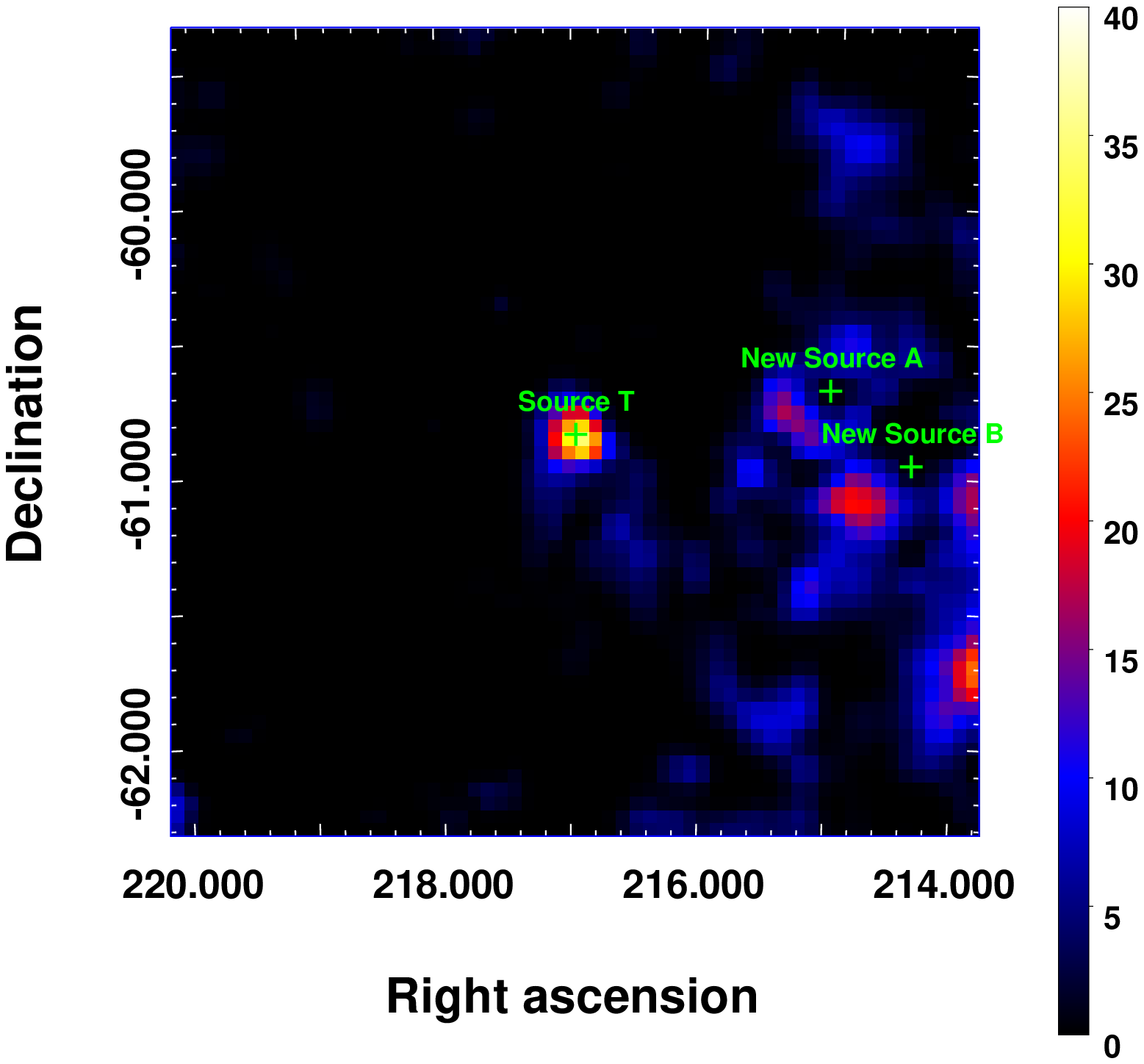}
\caption{TS maps of $3^{\circ}\times3^{\circ}$ region centered at HESS J1427-608 
for photons above 3 GeV. The left panel is for the standard model where contributions 
from 3FGL sources and diffuse backgrounds have been subtracted. The green crosses denote 
the newly added sources listed in Table \ref{table:newpts}. The right panel is the TS map with the additional new sources A and B subtracted. All maps are smoothed 
with a Gaussian kernel of $\sigma$ = $0.1^\circ$.}
\label{fig:tsmap}
\end{figure*}

The following analysis is performed using the latest Fermi-LAT Pass 8 data 
with {\it Source} event class from August 4, 2008 (Mission 
Elapsed Time 239557418) to November 4, 2015 (Mission Elapsed Time 468288004). 
Considering the impact of point spread function (PSF) and the complex diffuse 
emission from the Galactic plane, we only select events with energies between 
3 GeV and 500 GeV. Furthermore, in order to reduce the contamination from the 
Earth limb, the events with zenith angles larger than $90^\circ$ are excluded. 
The region of interest (ROI) we analyse is a $14^\circ \times 14^\circ$ square 
region centered at the position of HESS J1427-608 \citep[R.A.$=216.967^\circ$,
Dec.$=-60.85^\circ$;][]{Aharonian2008b}.
We analyse the data using the Fermi {\it ScienceTools} version {\tt
v10r0p5}\footnote{http://fermi.gsfc.nasa.gov/ssc/data/analysis/software/}
and the instrument response function (IRF) of P8R2{\_}SOURCE{\_}V6.
To model the Galactic diffuse emission and isotropic diffuse background,
{\tt gll\_iem\_v06.fits} and {\tt iso\_P8R2\_SOURCE\_V6\_v06.txt}\footnote
{http://fermi.gsfc.nasa.gov/ssc/data/access/lat/BackgroundModels.html}
are adopted. In addition to the diffuse backgrounds, 
the sources included in the 3FGL and located within the ROI are added to 
the model to account for $\gamma$-ray emission.
In the following analysis, the standard binned likelihood method with 
{\tt gtlike} is applied.

\subsection{Source detection}

In each {\tt gtlike} run, the normalizations and spectral parameters of sources
with distance smaller than $7^\circ$ to HESS J1427-608, 
as well as  the normalizations of the two diffuse backgrounds, are left free  
during the fitting. After including in the sky model the emission from 3FGL sources and the two diffuse backgrounds, we create a $3^\circ \times 3^\circ$ Test
Statistic (TS) map centered at HESS J1427-608 with {\tt gttsmap}, as shown in 
the left panel of Fig. \ref{fig:tsmap}. As can be seen from this TS map, there 
are still some significant excesses (marked by green crosses) that are not 
included in the model. Specially, there are some complicated excesses to the 
west of the ROI center, which we mark as New Source A and B. These excesses 
are spatially consistent with the two ``wings'' of the Kookaburra 
complex \citep[HESS J1420-607 \& HESS J1418-609;][]{Aharonian2006b}. 
The two ``wings" are two extended $\gamma$-ray sources and argued to be PWN 
candidates by \citet{Aharonian2006b}. \citet{Acero2013} explored high-energy 
$\gamma$-ray emission from these two sources with Fermi-LAT and concluded 
that HESS J1420-607 was a PWN candidate, and the emission from  HESS J1418-609 
was dominated by a pulsar. 
Besides the New Source A and B, we find an significant 
excess at the ROI center, the position of HESS J1427-608. This source, referred as 
Source T, is the focus of this work. Those three new sources are added in the model 
as point sources with power-law spectra. Their accurate positions are obtained with 
{\tt gtfindsrc} command. The resulting coordinates and TS values of them are listed 
in Table \ref{table:newpts}. 
The new TS map with sources A and B included in the model (and hence subtracted from the map) is shown in the right panel of Fig. \ref{fig:tsmap}. Except for Source T, this TS map is relatively clean with only small residuals.

\begin{center}
\begin{figure*}[!htb]
\centering
\includegraphics[width=0.8\textwidth]{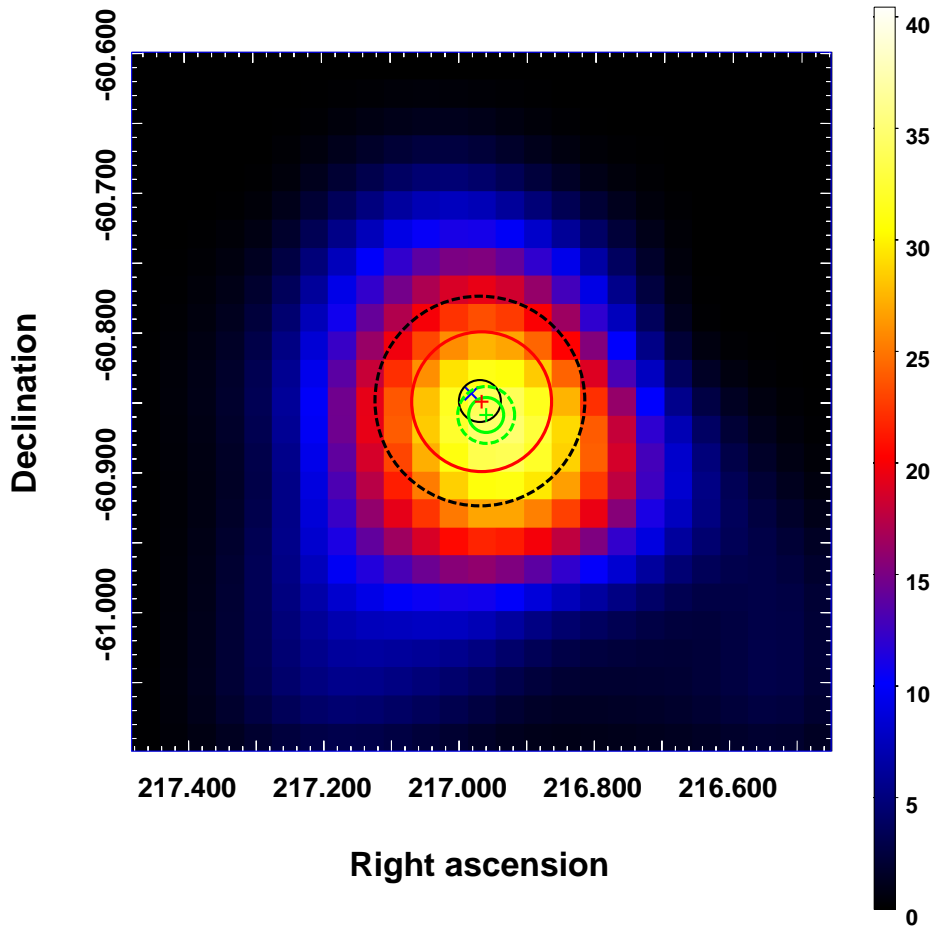}
\caption{TS map above 3 GeV for a region of $0.5^{\circ}\times0.5^{\circ}$
centered at the best-fit position of Source T. The pixel size is $0.02^{\circ}\times0.02^\circ$ 
and smoothing is applied with a Gaussian function of $\sigma=0.04^\circ$.
The green plus denotes the best-fitting position of Source T. The $1\sigma$ 
and $2\sigma$ positional error circles of Source T are marked by green solid 
and dashed circles, respectively. The position of HESS J1427-608 is marked 
as red plus and the red circle represents the extended morphology with radius 
of $\sigma=3^{'}$ \citep{Aharonian2008b}. The black solid circle shows the core 
region of Suzaku J1427-6051 with an extension of $\sigma=0.9^{'}$ and the source 
region with a radius of $4.5^{'}$ defined in \citet{Fujinaga2013} is indicated 
by the dashed one. A radio source named MGPS J142755-605038 in this region observed 
in the second epoch Molonglo Sky Survey \citep{Murphy2007} is marked by the blue cross.}
\label{fig:tsmap2}
\end{figure*}
\end{center}

Now we mainly discuss Source T. Source T is detected with a TS value 
of 40, which corresponds to a significance of $\sim5.5\sigma$ for four
degrees of freedom (dof). The best-fitting position of Source T is  
R.A.$=216.960^\circ$, Dec.$=-60.8595^\circ$ with $1\sigma$ error circle 
of $0.018^\circ$. The distance between Source T and the central position 
of HESS J1427-608 is only $0.7^{'}$, within the $1\sigma$ error circle 
of the Fermi-LAT data. To better understand the spatial correlation between 
Source T and HESS J1427-608, as well as candidate counterparts in other 
wavelengths, we create a zoom-in TS map for a region of 
$0.5^{\circ}\times0.5^{\circ}$ centered at Source T, as shown in Fig. \ref{fig:tsmap2}. 
The position and extended size of HESS J1427-608 \citep{Aharonian2008b}, 
as well as the X-ray counterpart Suzaku J1427-6051 \citep{Fujinaga2013}
are over-plotted. As we can see from Fig. \ref{fig:tsmap2}, Source T is 
well coincident with HESS J1427-608 and Suzaku J1427-6051. Therefore, 
Source T is very likely to be the GeV counterpart of HESS J1427-608.

\begin{table}[!htb]
\centering
\caption {Coordinates and TS values of the new point sources}
\begin{tabular}{cccc}
\hline \hline
Name & R.A. [deg] & Dec. [deg] & TS \\

\hline
Source T      & $216.960$ & $-60.8595$ & $40$\\
New Source A  & $215.028$ & $-60.6853$ & $63$ \\
New Source B  & $214.396$ & $-60.9555$ & $34$ \\
\hline
\hline
\end{tabular}
\label{table:newpts}
\end{table}

\subsection{Spatial extension}
Considering that HESS J1427-608 is slightly extended in the TeV and X-ray bands, 
we, therefore, re-perform the analysis with three uniform disks with different 
radii to test its spatial extension. The radii of the
three uniform disks we adopted are $0.1^\circ$, $0.15^\circ$ and $0.2^\circ$,
respectively. And the best-fitting TS values are 35, 31 and 24.
The larger the radius of the uniform disk is, the smaller the TS value is.
We also test the extension using the data with 4 PSF event-types\footnote
{http://fermi.gsfc.nasa.gov/ssc/data/analysis/documentation/Cicerone/Cicerone{\_}Data/LAT{\_}DP.html},
and find no significant spatial extension of the $\gamma$-ray emission. 
Therefore, a point-like source is enough to explain the $\gamma$-ray emission of 
Source T. In the following sections, we treat Source T as a point source.

\subsection{Spectral analysis}

\begin{figure}[!htb]
\centering
\includegraphics[angle=0,scale=0.35,width=0.5\textwidth,height=0.3\textheight]{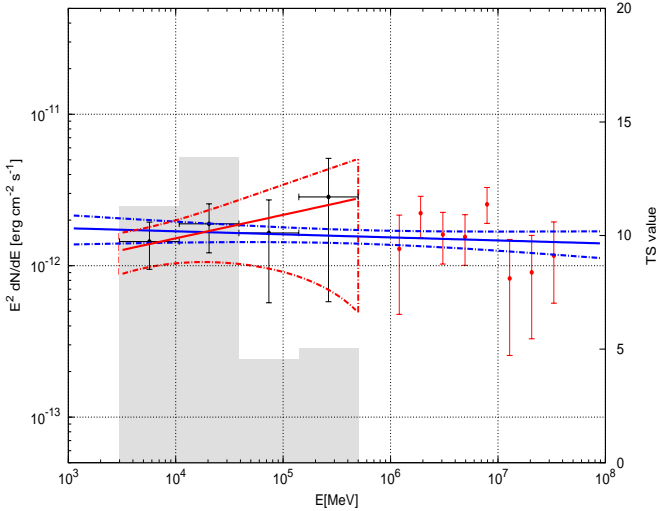}%
\hfill
\caption{SED of Source T. The black dots depict the Fermi-LAT data
in $3-500$ GeV and the high energy data observed by HESS are marked 
as red dots \citep{Aharonian2008b}. The gray histogram shows the TS 
value of each energy bin. The red solid and dotted-dashed lines indicate 
the best-fitting power law spectrum and $1\sigma$ statistical error for 
Fermi-LAT data in $3-500$ GeV, respectively. A power law spectra with an
index of $2.02\pm0.03$ for the global $\gamma$-ray data is plotted as 
the blue solid line, and its $1\sigma$ statistical error is also marked 
by the blue dotted-dashed lines.}
\label{fig:sed}
\end{figure}

To investigate the spectrum of Source T, we perform the global likelihood analysis
in the energy range between 3 GeV and 500 GeV. The spectrum of Source T
can be well described by a power law. The spectral index is $1.85\pm0.17$ 
and the integral photon flux is $(3.06\pm0.73)\times10^{-10}$ 
photon cm$^{-2}$ s$^{-1}$ with statistical error only. Since no information 
of the distance of HESS J1427-608 is available yet, we assume that d is 8 kpc 
following \citet{Fujinaga2013}. Thus the $\gamma$-ray luminosity above 3 GeV 
is $7.54\times 10^{34}\,(d/8\ {\rm kpc})^2$ erg~s$^{-1}$.

Then we bin the data into four logarithmically even energy bins from 3 GeV to 500 GeV, 
and repeat the likelihood fitting to give the SED. The spectral normalizations of 
all sources within $7^\circ$ from  Source T and the normalizations of the two 
diffuse backgrounds are set as free parameters, while the spectral indexes of 
these sources are fixed. The SED is plotted in Fig. \ref{fig:sed}. As we can 
see from the SED, the Fermi-LAT data connect with HESS's TeV spectrum smoothly,
which further supports that Source T is the GeV counterpart of HESS J1427-608.

\subsection{Timing analysis}

\begin{figure}[!htb]
\centering
\includegraphics[angle=0,scale=0.35,width=0.5\textwidth,height=0.3\textheight]{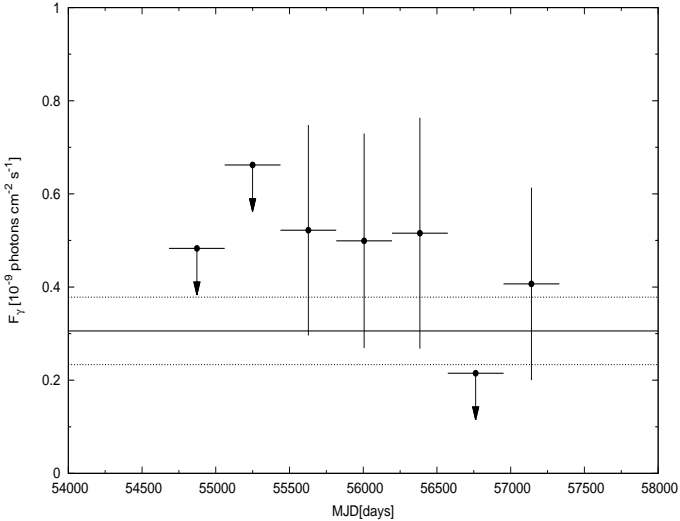}%
\hfill
\caption{Light curve for Source T. The horizontal solid line represents the average 
flux from the whole data set whose $1\sigma$ uncertainty is marked by the two 
dotted lines.}
\label{fig:lt}
\end{figure}

To check out whether Source T is variable or not, the timing analysis is preformed.
We bin the data into 7 equal time bins. The analysis procedure is the same as the 
spectral analysis. For any time bin whose TS value is smaller than 4, 95\% upper 
limit is calculated. The light curve is shown in Fig. \ref{fig:lt}. No obvious 
variability of Source T can be seen from this light curve. However, this may be 
partially due to the limited statistics of photons of the source. Further 
timing analysis with more data may be helpful to reveal its variable or steady nature.

\section{Discussion}

\subsection{Multi-wavelength observations}

From the above analysis, Source T is positionally consistent with 
HESS J1427-608 and its spectrum also connects with HESS observation 
smoothly, which all support that Source T is the GeV counterpart of
HESS J1427-608.

The X-ray emission observed by {\it Suzaku} is identified to be non-thermal 
\citep{Fujinaga2013}. And the unabsorbed flux in the 2-10 keV band is 
$\sim9 \times 10^{-13}~{\rm erg}~{\rm s}^{-1}~{\rm cm}^{-2}$ 
with the 90$\%$ upper limit in the 15-40 keV band of
$\sim5.3\times 10^{-12}{\rm erg}~{\rm s}^{-1}~{\rm cm}^{-2}$ \citep{Fujinaga2013}.

A radio compact source MGPS J142755-605038 with total flux density at 843 MHz 
of 34.5 mJy, discovered in the second epoch Molongo Sky Survey \citep[MGPS-2;][]{Murphy2007}, 
is located very close to HESS J1427-608. And \citet{Vorster2013} treated it as 
the radio counterpart of HESS J1427-608. 
However, the morphology of the radio source is distinct from that of TeV $\gamma$-rays.
\citet{Murphy2007} fitted the source morphology with a two-dimensional Gaussian function, and obtained that the full width at half maximum sizes of the major and minor axes ($\theta_M$, $\theta_m$) are 92.4$^{''}$ and 66.9$^{''}$, respectively. Such an extension is significantly smaller than that of the TeV $\gamma$-ray extension. Therefore we alternatively scale-up the flux density of MGPS J142755-605038 by a factor of 29 (the area ratio of the TeV $\gamma$-ray to the radio morphologies), as a very loose upper limit of the radio emission of HESS J1427-608. The actual upper limit should be lower than such a result.


\subsection{Radiation mechanism}

Here we consider two scenarios for the origin of $\gamma$-ray emission.
In the leptonic scenario, the $\gamma$-ray emission can be produced by
inverse Compton scattering (ICS) or bremsstrahlung process of high 
energy electrons. In the hadronic scenario, the $\gamma$-ray photons 
originate from the decay of neutral pion mesons produced in collisions 
between accelerated protons and the ambient materials. The Fermi-LAT 
results we get and the TeV $\gamma$-ray data observed by HESS will 
be used to study the multi-wavelength SED of HESS J1427-608.

In the following, we quantitatively discuss the leptonic and hadronic scenarios
with the multi-wavelength data. In the modelling, a power law spectrum with 
an exponential cutoff for electrons and protons is assumed: 
$dN/dE_i \propto E_i^{-\alpha_i} \exp(-E_i/E_{i,\rm cut})$,
where $i = e, p$, $\alpha_i$ is the spectral index and $E_{i, \rm cut}$ is 
the cutoff energy. Besides the cosmic microwave background (CMB), the 
radiation field also includes an infrared blackbody component 
($T_{\rm IR}=30$ K, $U_{\rm IR}=1$ eV cm$^{-3}$) and an optical blackbody 
component ($T_{\rm opt}=6000$ K, $U_{\rm opt}=1$ eV cm$^{-3}$) \citep{Porter2006}.
The distance of HESS J1427-608 is taken to be 8 kpc, which follows the 
assumption in \citet{Fujinaga2013}. And the radius of HESS J1427-608 
is $\sim$ 7 pc correspondingly. Here the gas density of 
$n_{\rm gas} = 1$ cm$^{-3}$ is assumed.

\subsubsection{Leptonic scenario}

\begin{center}
\begin{figure*}[htbp]
\centering
\includegraphics[height=2.3in]{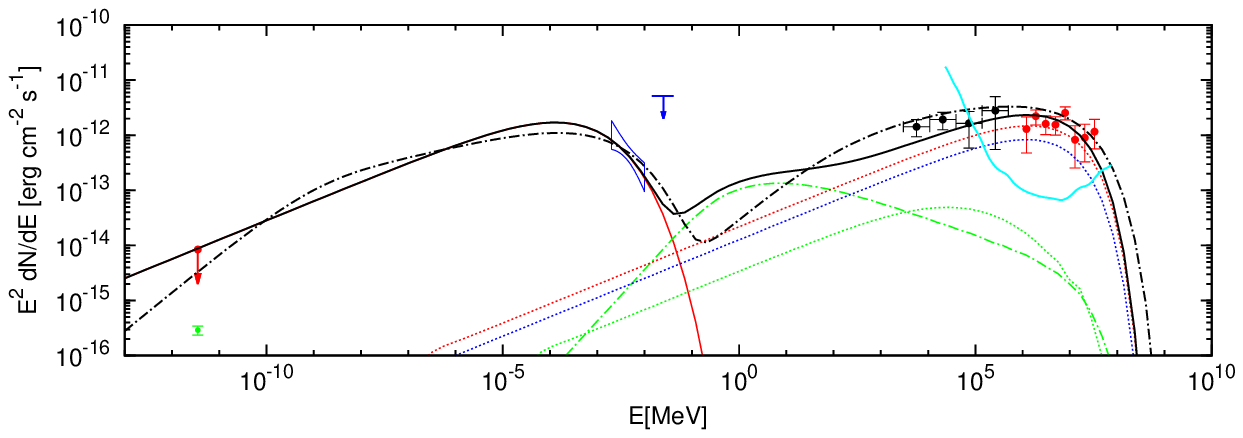} 
\includegraphics[height=2.3in]{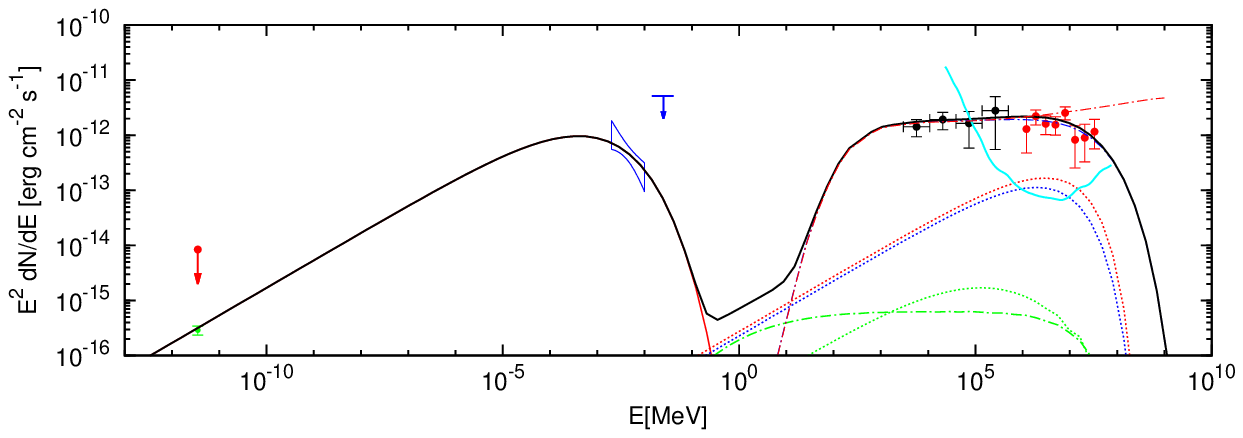} 
\caption{Multi-wavelength SED of HESS J1427-608. 
The green point is the radio flux of MGPS J142755-605038 \citep{Murphy2007}
and the red arrow shows the upper limit recalculated based on the size of 
MGPS J142755-605038 and HESS J1427-608.
The X-ray data marked by the 
blue butterfly and arrow are taken from {\it Suzaku} observation \citep{Fujinaga2013}.
The top panel is for the leptonic scenario and the bottom one is for the hadronic 
scenario. The radio and X-ray emissions are dominated by a synchrotron component, 
shown as a red solid curve. The $\gamma$-ray emission is mainly modelled by different 
contributions, including $\pi^0$ decay (blue dotted–dashed), bremsstrahlung 
(green dotted–dashed) and ICS (dotted) components. And the ICS includes three 
components from CMB (blue), infrared (red), and optical (green) radiation fields. 
The black solid line represents the sum of different radiation components.
The cyan curve shows the differential sensitivity of CTA \citep[100 hours;][]{Funk2013}. 
The black dotted-dashed line in the top panel denotes the sum of different radiation 
components produced by the electrons with a broken power law distribution and
the red dotted-dashed line in the bottom panel represents the $\pi^0$ decay 
component for a proton distribution without a high-energy cutoff.}
\label{fig:multi-sed}
\end{figure*}
\end{center}

For the leptonic scenario, all these fitted parameters are listed in 
Table \ref{table:model}, which are used to calculate the radiation spectrum
as shown in the top panel of Fig. \ref{fig:multi-sed}. To explain the emission from 
radio to TeV band, we get the model parameters of $\alpha_{\rm e} \approx 2.3$ 
and $E_{e, \rm cut} \approx 35~{\rm TeV}$. While the radio flux upper limit favors 
a lower value of $\alpha_e$, the $\gamma$-ray spectrum prefers a value of 3. 
A broken power-law model will lead to a better fit to the broadband SED,
as shown in the top panel of Fig. \ref{fig:multi-sed}.
However, if MGPS J142755-605038 is treated as the radio counterpart of HESS J1427-608, 
the spectra of electrons should be harder, which is difficult to explain the flat spectrum 
in the $\gamma$-ray band. This will be a challenge to the leptonic scenario.

A very low strength of magnetic field, $B\approx 5.0$ $\mu$G, is derived in order 
not to overproduce the synchrotron emission in the radio band in this scenario. 
This value is very close to the strength of Galactic magnetic field. The total 
energy of electrons above 1 GeV is estimated to be 
$W_{\rm e}\approx 2.5 \times 10^{48} (d/8.0\,\mathrm{kpc})^{2}\ \mathrm{erg}$, 
which would vary with the value of distance adopted.

\begin{table*}
\centering
\normalsize
\caption {Parameters for the models}
\begin{tabular}{cccccccccccccc}
\hline \hline
Model & $\alpha_p$  & $\alpha_e$  & $\Delta \alpha_e$  & $E_{p,\rm cut}$  & $E_{e, \rm break}$  & $E_{e, \rm cut}$  & $W_p$  & $W_e$  & $B$  & $n_{\rm gas}$\\
      &     &     &      & (TeV)     & (TeV)  & (TeV)   & ($10^{50}$ erg) & ($10^{48}$ erg) & ($\mu$G) & (cm$^{-3}$)\\
\hline
leptonic (PL)  & $-$   & $2.3$  & $-$  & $-$   & $-$   & $35.0$  & $-$   & $2.5$ & $5.0$ & $1.0$ \\
\hline
leptonic (BPL) & $-$   & $1.6$  & $1.0$  & $-$   & $0.1$   & $80.0$  & $-$   & $3.0$ & $3.5$ & $1.0$ \\
\hline
hadronic  & $2.0$ & $2.0$  & $-$  & $350.0$  & $-$  & $30.0$ & $10.0$  & $0.06$ & $10.0$ & $1.0$ \\
\hline \hline
\end{tabular}
\label{table:model}
\tablecomments{The total energy of relativistic particles, $W_{e,p}$, is
calculated for $E > 1$ GeV.}
\end{table*}


\subsubsection{Hadronic scenario}

The fitted multi-wavelength SED with hadronic scenario is compiled in 
the bottom panel of Fig. \ref{fig:multi-sed} and the model parameters 
are listed in Table \ref{table:model}.

In this model, the spectral indices of protons and electrons are set 
equal to 2.0 in view of the diffusive shock acceleration mechanism 
and the magnetic field strength is adopted to be a typical value of 
$B\approx 10$ $\mu$G to reduce the number of free parameters in the model. 
Based on this, a cutoff energy of $\sim 30$ TeV and total energy above 
1 GeV of $W_{\rm e}\approx 0.6 \times 10^{47} (d/8.0\,\mathrm{kpc})^{2}\ \mathrm{erg}$ 
for electrons are needed to explain the flux in radio and X-ray band. 
The total energy of protons above 1 GeV is 
$W_p \approx 1.0 \times 10^{51} (n/1.0\,\mathrm{cm}^{-3})^{-1}\ (d/8.0\,\mathrm{kpc})^{2}\ \mathrm{erg}$. 
However, this parameter strongly depends on the values of gas density and 
distance adopted here.
Since there is no evident cutoff in the photon spectrum 
in the TeV energy band, the cutoff energy of protons should be larger than 
$\sim 350$ TeV for the $\gamma$-ray spectra with an index of 2.0 obtained above. For comparison, 
the $\pi^0$ decay component for protons without a high-energy cutoff is 
plotted in the bottom panel of Fig. \ref{table:model}. 
HESS J1427-608 is a good candidate to accelerate CR protons up 
to several hundreds TeV or even PeV energies. This is a good opportunity for the 
Cherenkov Telescope Array (CTA) to test the PeVatron nature of HESS J1427-608.
Note that for a distance of 8 kpc, the pair production absorption of the VHE 
photons by the infrared and/or CMB photons is non-negligible \citep{Zhang2006}, 
and the extension of the power-law spectrum to PeV energies should be treated as an upper limit. 

As can be seen in Fig. \ref{fig:multi-sed}, both the leptonic and hadronic models
can fit the multi-wavelength data of HESS J1427-608.

\subsection{The possible nature of HESS J1427-608}

\begin{figure}[!htb]
\centering
\includegraphics[angle=0,scale=0.35,width=0.5\textwidth,height=0.3\textheight]{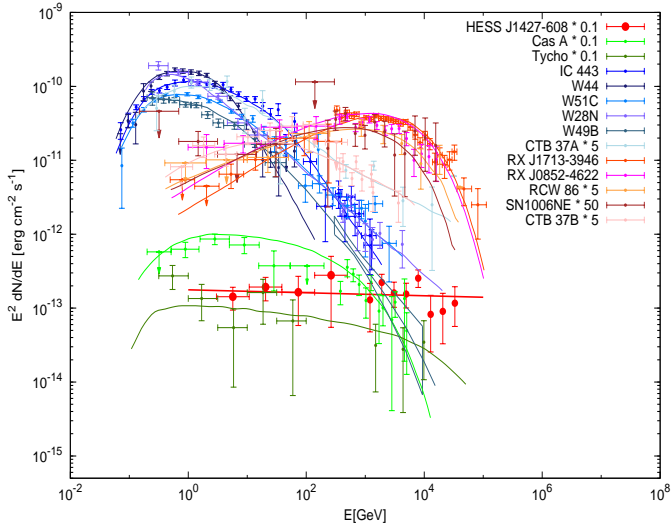}%
\hfill
\caption{Typical $\gamma$-ray SED for several of the most prominent SNRs and HESS J1427-608.
The energy fluxes of several sources with relatively flat spectra are scaled downward 
by one order of magnitude for the sake of clarity.}
\label{fig:SNR-sed}
\end{figure}

Considering that most of the Galactic VHE sources are identified to be 
PWNe or SNRs, here we discuss these two possibilities as the origin of
HESS J1427-608. 

\subsubsection{SNR}

First, we discuss the possibility that the emission comes from a SNR. 
For comparing between HESS J1427-608 and several of the most prominent
SNRs, we plot the $\gamma$-ray SED of them in Fig. \ref{fig:SNR-sed}.
The energy fluxes of several SNRs are adjusted for the sake of illustration \citep{Funk2015}, 
as shown in the legend of Fig. \ref{fig:SNR-sed}. As can be seen, these SNRs
can be divided into three classes. The first class is known to be interacting
with molecular clouds, such as IC 443 \citep{Abdo2009, Ackermann2013}, 
W44 \citep{Abdo2010c, Ackermann2013}, W51C \citep{Abdo2010f, Jogler2016}, 
W28 \citep{Abdo2010g, Aharonian2008c}, W49B \citep{Abdo2010h, Brun2011} 
and CTB 37A \citep{Brandt2013, Zeng2016}. These SNRs are old and 
typically brighter in the GeV than in the TeV band. And the $\gamma$-ray 
emissions of these sources are considered to be from hadronic process. 
Especially, IC 443, W44 and W51C have shown the characteristic 
``$\pi^0$ bump'' \citep{Ackermann2013, Jogler2016}, which are considered 
to be the most direct evidence that protons are accelerated in SNRs. 
Differing from this class, the second class, such as 
RX J1713-3946 \citep{Abdo2011, Yuan2011, Zeng2016}, RX J0852-4622 \citep{Tanaka2011}, 
RCW 86 \citep{Yuan2014}, SN1006 \citep{Acero2010, Araya2012, Xing2016}
and CTB 37B \citep{Xin2016, Zeng2016}, has harder spectra in the GeV band 
and is brighter in the TeV band than in the GeV band. The sources in this 
class are middle-aged SNRs and generally thought to have the leptonic origin 
for the $\gamma$-ray emission although several of them are still being debated. 
The third class in Fig. \ref{fig:SNR-sed} includes Cas A \citep{Abdo2010b, Yuan2011} 
and Tycho \citep{Giordano2012, Morlino2012, Zhang2013}, which are the youngest SNRs. 
And the $\gamma$-ray spectra of these two SNRs can be fitted by a power law extending 
from the GeV to TeV band, which can be well explained as the hadronic emission. 
Like Cas A and Tycho, HESS J1427-608 also has a flat $\gamma$-ray spectra from 
GeV to a few tens of TeV and can be fitted well by a power law with an index of 2.0.

For the leptonic scenario of HESS J1427-608, we compare the fitted parameters 
in Table \ref{table:model} with those the second class mentioned above in Fig. \ref{fig:SNR-sed}.
And we find that the total energies of electrons of these SNRs are in the order of 
$\sim 10^{47}-10^{48}$ erg and the fitting result of HESS J1427-608 is also 
in coincidence with that under the assumption of distance to be 8 kpc.
However, the fitting magnetic field strength of HESS J1427-608 is at least 
two times lower than those of other SNRs, which may be attributed to 
variations in the progenitor and its environment. However, the X-ray flux 
is dominated by a central bright source instead of a shell structure of 
other SNRs. The model predicts a sharp cutoff beyond the HESS energy range 
which can be tested with future CTA observations.

For the hadronic scenario, parameters in our model are compared with SNRs 
whose $\gamma$-ray emissions are expected to be dominated by $\pi^0$ decay.
The total energy of protons of HESS J1427-608,
$W_p \approx 1 \times 10^{51} (n/1.0\,\mathrm{cm}^{-3})^{-1}\ (d/8.0\,\mathrm{kpc})^{2}\ \mathrm{erg}$, 
is significantly higher than the typical energy of particles, assuming that
$\sim 10 \%$ of the supernova kinetic energy of $\sim 10^{51}$ erg is transferred to
the energy of particles. On the one hand, this source may be left from a hypernova
explosion which releases a substantially higher energy than typical supernova.
On the other hand, much higher value of $W_p$ depending on the gas density and the
distance means that HESS J1427-608 may be closer or the ambient environment 
of this source may be denser. Comparing with the first class SNRs, also known 
as SNRs interacting with molecular clouds leading to prominent thermal X-ray emissions,
the X-ray emission from HESS J1427-608 is dominant by non-thermal process. The absence 
of shell structure in the X-ray band for HESS J1427-608 and no adjacent SNRs or molecular clouds founded 
around make it distinct from others although several dark clouds are reported near the 
direction of HESS J1427-608 in the SIMBAD database \footnote{http://simbad.u-strasbg.fr/simbad/}. 
In addition, the flat $\gamma$-ray spectra of HESS J1427-608 from GeV to a few tens 
of TeV is also significantly different from that of the first class SNRs but
similar to that of the third class SNRs, especially Tycho, which also
emit the non-thermal X-ray emissions. For HESS J1427-608, the fitting cutoff 
energy of $\sim350$ TeV or even 1 PeV for protons is about one order of magnitude 
higher than the values of most SNRs, which suggests that HESS J1427-608 is a promising 
PeV CR accelerator (PeVatron). However the absence of shell structure in the X-ray band 
and the weak radio emission are also challenging to this possible origin.

\subsubsection{PWN}

The $\gamma$-ray emission from PWNe is dominated by the ICS 
process \citep {Abdo2010a, Abdo2010d, Abdo2010e}.
\citet {Fujinaga2013} discussed the PWN hypothesis of HESS J1427-608 with
the X-ray and TeV data, and indicated that a simple one-zone model with 
magnetic filed strength of about 5 $\mu$G would roughly explain the SED, 
and our result is consistent with that. Our GeV data, in combination with 
the flux upper limit in the radio band, however, favors a more complex 
broken power-law model. In addition, \citet {Vorster2013} 
also modelled the multi-wavelength data of HESS J1427-608 as a PWN.
The fitting parameters of the leptonic scenario listed in Table \ref{table:model} 
are comparable with that of PWN Vela-X \citep {Abdo2010d}.
 
The X-ray, GeV and TeV luminosities of HESS J1427-608 are 
$L_{\rm X}(2 - 10{\rm keV})$ = $7 \times 10^{33}\,(d/8\ {\rm kpc})^2$ erg s$^{-1}$,
$L_{\rm GeV}(> 10{\rm GeV})$ = $5.21 \times 10^{34}\,(d/8\ {\rm kpc})^2$ erg s$^{-1}$, 
and $L_{\rm TeV} (1 - 30{\rm TeV})$ = $4.18 \times 10^{34}\,(d/8\ {\rm kpc})^2$ erg s$^{-1}$ \citep {Aharonian2008b},  
respectively. These estimated luminosities are within the value range of known 
PWNe or PWN candidates which have emissions in the X-ray and $\gamma$-ray bands \citep {Acero2013, Mattana2009}.
The ratio between the X-ray and $\gamma$-ray luminosities is approximately to be 1, which
is consistent with the value of magnetic field strength in our fitting and the value 
of about 5 $\mu$G is also typical for PWNe. \citet {Mattana2009} and \citet {Acero2013} 
studied a large sample of PWNe and derived the relations among $F_{\rm TeV}/F_{\rm X}$, $L_{\rm GeV}/L_{\rm X}$, 
spin-down energy \.E and characteristic age $\tau_c$ of the pulsars. 
The ratio between the GeV and X-ray luminosities of HESS J1427-608 
is estimated to be about 7.44, and 
according to these relations in \citet {Mattana2009} and \citet {Acero2013}, 
the value of \.E and $\tau_c$ of a pulsar to power HESS J1427-608 are suggested to 
be $\sim 6.5 \times 10^{36} $ erg s$^{-1}$ and 11 kyr, respectively. The values 
of \.E and $\tau_c$ are typical for Vela-like pulsars.

For the PWN hypothesis, the discussion above indicates that there may exist 
a Vela-like pulsar in the vicinity of HESS J1427-608. The detection of a pulsar 
near HESS J1427-608 will strongly support such a scenario.
Therefore, the future observations, especially the pulsation search in the
radio and other energy bands, are of great importance to address this hypothesis.

\section{Conclusion}

We investigate more than seven years data of the Fermi-LAT in the direction 
of the unidentified TeV source HESS J1427-608. A point source is detected with a 
significance of $\sim5.5\sigma$. This GeV source 
is positionally coincident with HESS J1427-608, and the GeV spectrum can 
connect with the TeV spectrum smoothly. These facts  suggest that 
this source is very likely the GeV counterpart of HESS J1427-608. 

We collect multi-wavelength data to constrain the radiation model of HESS J1427-608.
Both the leptonic and hadronic scenarios can fit the SED. At present it is difficult to 
exclude either one considering the large uncertainties of the data. A remarkable spectral 
feature of this source is its $E^{-2}$ spectrum over four decades of energies without 
obvious cutoff. This is unique among all currently detected $\gamma$-ray sources. 
If CR protons are responsible for the $\gamma$-ray emission, the highest energy of 
protons should exceed a few hundred TeV and even PeV. This makes HESS J1427-608 a 
promising PeV CR accelerator (PeVatron).

Unfortunately, the nature of HESS J1427-608 is not clear yet. Either a SNR or a PWN 
seems to be plausible. The low strength of magnetic field and the absence of shell 
structure in the X-ray band indicate that HESS J1427-608 may be a PWN rather than 
a SNR. However, its flat $\gamma$-ray spectrum in a wide energy range is more close 
to some SNRs but different from typical PWNe. Further multi-wavelength observations, 
in the radio and VHE $\gamma$-rays, are crucial to understand its nature ultimately. 
Especially the precise measurements of the $\gamma$-ray energy spectrum up to hundreds 
of TeV energies by, e.g., the CTA, will crucially test its PeVatron nature.

\section*{Acknowledgments}
This work is supported by the Natural Science Foundation of 
Jiangsu Province of China (No. BK2014102SBZ0140), 
the 973 Programme of China (No. 2013CB837000), the National Natural Science Foundation
of China (Nos. 11433009, 11233001 and 11303098), the Strategic Priority Research Program, the Emergence of Cosmological Structures, of the Chinese Academy 
(No. XDB09000000), and the 100 Talents program of Chinese Academy of Sciences.

\end{document}